\documentclass[twocolumn,preprintnumbers,aps,prd,superscriptaddress,nofootinbib,10pt,floatfix]{revtex4-2}
\usepackage{amssymb}
\usepackage{amsmath}
\usepackage{bbm}
\usepackage{hyperref}
\usepackage{graphicx}
\usepackage{orcidlink}
\usepackage{microtype}
\usepackage{slashed}

\newcommand{\commutator}[2]{\left[#1,#2\right]}

\usepackage{fbox}

\begin{document}
\preprint{TUM-EFT 205/26}

\title{Chiral symmetry restoration effects onto the meson spectrum from a Dyson-Schwinger and Bethe-Salpeter approach}

\author{Reinhard Alkofer\,\orcidlink{0000-0001-7433-3295}}
\email{reinhard.alkofer@uni-graz.at}
\affiliation{Institute of Physics, University of Graz, NAWI Graz, Universit\"atsplatz 5, 8010 Graz, Austria}

\author{Christian S. Fischer\,\orcidlink{0000-0001-8780-7031}}
\email{christian.fischer@theo.physik.uni-giessen.de}
\affiliation{Institut für Theoretische Physik, Justus-Liebig-Universität Gießen, Heinrich-Buff-Ring 16, 35392 Gießen, Germany}
\affiliation{Helmholtz Forschungsakademie Hessen für FAIR (HFHF), GSI Helmholtzzentrum für Schwerionenforschung, Campus Gießen, Heinrich-Buff-Ring 16, 35392 Gießen, Germany}

\author{Fabian Zierler\,\orcidlink{0000-0002-8670-4054}}
\email{fabian.zierler@tum.de}
\affiliation{Technical University of Munich, TUM School of Natural Sciences, Physics Department, James-Franck-Str. 1, 85748 Garching, Germany}
\affiliation{Centre for Quantum Fields and Gravity, Faculty  of Science and Engineering, Swansea University, Singleton Park, SA2 8PP, Swansea, United Kingdom}
\affiliation{Department of Physics, Faculty of Science and Engineering, Swansea University, Singleton Park, SA2 8PP Swansea, Wales, United Kingdom}

\begin{abstract}
Light meson spectra are studied in a Dyson-Schwinger/Bethe-Salpeter approach to QCD.
By varying the interaction strength of three sets of models for the quark-antiquark 
interaction, the transition from the chiral symmetric to the chirally broken regime 
in the vacuum is studied. The simplest type of these models leads to degenerate meson 
spectra for a large domain of the strength parameter. The more sophisticated and thus 
more realistic models show significantly smaller parameter domains for which degenerate 
meson spectra are obtained. The underlying mechanism for obtaining and then lifting 
degeneracies is traced back to the location of the quark propagators' poles, in 
particular, whether they are beyond or within the domain of integration in the 
Bethe-Salpeter equation. In view of this mechanism the potential relation of the 
obtained degeneracies to the dynamical emergence of symmetries is discussed, 
adding thereby another point of view on the conjectured chiral spin symmetry of QCD 
in the temperature domain right above the crossover. 
\end{abstract}

\maketitle
\section{Introduction}\label{sec:intro}

Degeneracies of states are typically related to symmetries. Hereby the
corresponding multiplets are used together with the quantum numbers
of the states to infer the underlying symmetry transformations. 
A recent example is given by the observation of meson degeneracies in
QCD lattice calculations at temperatures just above the crossover
temperature $T_\chi$ \cite{Rohrhofer:2019qwq} and their interpretation 
in terms of the so-called chiral spin symmetry, see 
\cite{Glozman:2022lda,Glozman:2022zpy} and references therein. 

This emergent symmetry raises immediately questions about its nature,
first of all,  because it goes beyond the symmetry of the free 
Dirac Lagrangian. A possible 
interpretation might be linked to the fact that the chiral spin symmetry 
is a symmetry of the purely chromoelectric part of QCD, for respective 
details see, e.g., \cite{Glozman:2022zpy}, where it is also suggested
that in a temperature range from $T_\chi$ to $\sim 3 T_\chi$
strongly-interacting matter forms a fluid of hadrons with still quite
strong interactions in between them.

Although the current knowledge indicates that chiral spin symmetry 
smoothly disappears above $\sim 3 T_\chi$ it is noteworthy in this
context that quite recently there has been found some evidence that
QCD undergoes two cross-overs, resp., phase transitions  
\cite{Mickley:2024vkm} with the second transition being related 
to the percolation properties of center vortices and thus to 
confinement. Hereby the higher transition temperature at $\sim 2 T_\chi$
is not so far from the estimate for the temperature at which chiral
spin symmetry ceases to be realized. In a very recently suggested 
related scenario \cite{Fujimoto:2025sxx} arguments for an intermediate 
phase of strongly-interacting matter in between the hadronic and the 
high-temperature phases have been provided. In this phase the thermal
degrees of freedom of quarks behave as deconfined whereas the gluons
are still confined within glueballs. 

Furthermore, based on spatial and temporal lattice correlators
as well as theoretical constraints like micro-causality, the KMS
condition, etc., the persistence of resonance-like structures 
in the pseudo-scalar channel also above $T_\chi$ has been demonstrated
\cite{Lowdon:2024fgv}. Hereby the authors of this study state 
explicitly that this is consistent with predictions based on an 
emergent chiral spin symmetry.

Given this situation a deeper insight into the nature of this emergent
chiral spin symmetry is certainly desirable.

The starting point of our investigation is hereby the observation 
of degeneracies of meson spectra, or more precisely, how can 
degeneracies be obtained when taking the nature of mesons as 
quark-antiquark bound states into account. Are there 
degenerate spectra of mesons for certain types of quark-antiquark
interactions? 
How do degeneracies depend on the interaction strength? Which quantum numbers are involved?
The latter question is particularly interesting. Traditionally, the focus has been on chiral 
partners such as $\pi,\sigma$ and $\rho, a_1$. The degeneracy of the latter has been flagged 
as an important signal for chiral symmetry restoration which can be extracted from the dilepton
spectrum of a heavy-ion collision\cite{Rapp:1999ej}. Chiral spin symmetry, however, involves
a much larger set of degeneracies \cite{Glozman:2022zpy}. 

To contribute to the answers to these questions we will employ a Dyson-Schwinger/Bethe-Salpeter (DS/BS) 
approach and investigate the resulting meson bound states based on different interaction models. For
simplicity, we restrict ourselves to the vacuum for the moment and defer a study at finite temperature
to later work. The role of the symmetry restoring parameter is then shifted from temperature
to corresponding model parameters for the interaction strength. While this setup is certainly not 
elaborate enough to allow for definite conclusions on the potential high temperature phase discussed
in \cite{Glozman:2022zpy}, it is interesting in itself and allows for the identification of potential 
underlying dynamical mechanisms beyond simple symmetry considerations. Quite surprisingly a full 
pattern of extended and partial degeneracies can be found. Even more astonishingly, these patterns 
start (or better end) at parameter values for which the quark propagators' poles enter the 
integration domain for the integrals appearing in the BS equation

This paper is organized as follows: In the next section we present 
three interaction models to be used first in the DS equation for 
the quark propagator. In Sect.\ \ref{sec:BSE}
we briefly review the BS equation 
for pseudo-scalar, vector, scalar and axial-vector mesons including the
implementation of the three model interactions. 
In Sect.\ \ref{sec:results} we present numerical results, first, for the behavior of the 
quark propagators' low-lying poles, and second, for the resulting 
spectra of mesons. We will identify the occurrence and the lifting of 
degeneracies as a function of the used parameter sets and relate them 
to the locations of the quark propagators' poles. In the concluding 
section we will discuss the mechanism uncovered by our investigation 
as well its implication for the emergent chiral spin symmetry. 
In the appendix we provide several additional plots for the 
location of the quark propagator's poles. 

\section{Dyson-Schwinger Equation and Models}\label{sec:models}

We work in Euclidean spacetime. In this case bound states occur at $-M^2 = P^2 < 0$ and the BS equations used to calculate meson 
masses probe the quark propagator in a parabolic region of the complex plane, see e.g. \cite{Eichmann:2016yit} for an overview. 
The quark DS equation reads:
\begin{align}
S^{-1}(p) = S_0^{-1}(p) + Z_{1F} C_F g^2 \int_q D^{\mu\nu}(k) \gamma^\mu S(q) \Gamma^\nu(q,p) \, .
\end{align}
Here $S(p)$ denotes the quark propagator for quark momentum $p$, $\Gamma^a_\nu(q,p)$ the quark gluon vertex for momenta $q$ and $p$ and $D_{\mu\nu}(k=p-q)$ the gluon propagator. The notation $\int_q$ is a short form of $\int \text{d}^4 q / (2\pi)^4$.
$S_0(p)$ is the bare quark propagator, $Z_{1F}$ the multiplicative renormalization constant of the bare quark-gluon vertex,  
and $C_F$ the Casimir of the fundamental representation taking the value $4/3$ for SU(3).

The quark propagator's Dirac tensor structure can be decomposed into two terms, and thus the quark propagator as well as its inverse are described by two scalar dressing functions:
\begin{align}
  S^{-1}(p)&=+i\gamma_\mu p^\mu A(p^2)+\mathbbm{1}B(p^{2}) \, , \\
  S(p)&=-i\gamma_\mu p^\mu \sigma_A(p^2)+\mathbbm{1}\sigma_B(p^{2}) \, .
\end{align}
The dressing functions of the inverse quark propagator are related to those of the quark propagator by
{}
\begin{align}
    \sigma_A(p^2) = \frac{A(p^2)}{p^2 A^2(p^2) + B^2(p^2)} \equiv  A(p^2) \sigma(p^2), \label{sigmaA} \\ 
    \sigma_B(p^2) = \frac{B(p^2)}{p^2 A^2(p^2) + B^2(p^2)}
    \equiv  B(p^2) \sigma(p^2). \label{sigmaB}
\end{align}
The quark mass function is then given by
\begin{equation}
    M(p^2) = B(p^2) / A(p^2) = \sigma_B(p^2) / \sigma_A (p^2) \, .
\end{equation}

In the following we briefly describe the three interaction models studied in this work. Our choice is guided by considerations of 
simplicity and feasibility. However, all three model types have already been tested extensively in their capability to describe 
the QCD phase diagram and found to be qualitatively useful, see \cite{Fischer:2018sdj} for a review. Model III can even be seen 
as a (slightly simplified) vacuum version of semi-quantitative successful truncations that are employed to predict the location 
of the critical endpoint of QCD \cite{Fischer:2014ata}. Thus while our study cannot anticipate more involved investigations 
using the most advanced truncation schemes at finite temperature and chemical potential available today
\cite{Fu:2019hdw,Gao:2020fbl,Gunkel:2021oya} it serves as a starting point with the potential to add qualitative insights to 
the discussion on chiral spin symmetry. 

\subsection{Model I}

The first and simplest model of interest is the interaction proposed in Ref.~\cite{Alkofer:2002bp}. 
It is Gau\ss ian-like suppressed in the ultraviolet momentum regime, but provides sufficient strength at small momenta
to trigger dynamical chiral symmetry breaking:
\begin{align}
	\label{eq:alkofer-watson-weigel}
	Z_{1F}g^2 D^{\mu\nu}(k)\Gamma^\nu(q,p) & \to Z_2^2 \, \mathcal{G}\left(k^2\right) D_{free}^{\mu\nu}(k)\gamma^\nu \, , \\
	\mathcal{G}(k^2) &= 4\pi^{2} \frac{D_{I}}{\omega_I^{6}} k^{4}e^{-\frac{k^{2}}{\omega_I^{2}}} \, .
\end{align}
Here, $Z_2$ is the renormalization constant of the quark propagator and 
$D_{free}^{\mu\nu}(k) = \left( \delta_{\mu \nu} - \frac{k_\mu k_\nu}{k^2} \right) \frac{1}{k^2}$.
The parameter $D_I$ has mass dimension two and $\omega_I$ has 
mass dimension one. $\mathcal{G}$ stands for a dimensionless effective running coupling. 
Both parameters are associated with chiral symmetry breaking: $D_I/\omega_I^6$ is a measure for the interaction strength, while $\omega_I$ is associated with a scale. {
In our study below we decrease both parameters simultaneously while keeping the ratio $D_I/\omega_I^6=64$ {GeV$^{-4}$} constant, see table \ref{tab:param_W} in appendix \ref{sec:appendix_more_plots}.
The initial values $D_I=1$ {GeV$^2$} and $\omega_I=0.5$ {GeV} are taken from Ref.~\cite{Alkofer:2002bp} and correspond to their $D=16$ {GeV$^{-2}$} and $\omega=0.5$ {GeV} at the physical point of the model. 
}

{In Sec.~\ref{sec:results} we present results for the quark propagators' poles in the chiral limit and for a small current quark mass 
$m[\mu = 19\, \mathrm{GeV}]=0.005$~GeV in a momentum subtraction scheme. 
All meson masses are calculated with the non-vanishing current quark mass.}

For this interaction the quark propagator DSEs can be solved in the entire complex plane at high precision. We follow the approach of Ref.~\cite{Windisch:2016iud} and use pole-counting integrals to identify the pole locations in the complex plane. We remark that our results agree well with the ones reported in \cite{Dorkin:2013rsa,Windisch:2016iud} for this interaction. 

\subsection{Model II}

A phenomenologically successful effective interaction for the light pseudo-scalar and vector mesons that also captures the correct ultraviolet behaviour of an effective running coupling is given by \cite{Maris:1999nt}:
\begin{align}
  \label{eq:maris-tandy}
  \mathcal{G}(k^2) &= \mathcal{G}_{\rm IR}(k^2) + \mathcal{G}_{\rm UV}(k^2), \\ 
   \label{eq:maris-tandyIT}
  \mathcal{G}_{\rm IR}(k^2) &= 4\pi^{2}\frac{D_{II}}{\omega_{II}^{6}} k^{4}e^{-\frac{k^{2}}{\omega_{II}^{2}}},  \\  
   \label{eq:maris-tandyUV}
  \mathcal{G}_{\rm UV}(k^2) &= \frac{4\pi^{2} \gamma_m \left( 1 - \exp\left[-\frac{k^2}{4m_t}\right] \right) }{\frac{1}{2} k^2 \ln\left[ \tau +\left(1+\frac{k^{2}}{\Lambda_{QCD}^{2}}\right)^{2}\right]}. 
\end{align}
The infrared term is identical to Eq.~\eqref{eq:alkofer-watson-weigel}, while the ultraviolet term ensures the correct perturbative running of the solutions for the quark propagator. Hereby, $\gamma_m$ is the anomalous dimension of the quark mass. For 
the choice of the technical parameters $m_t$, $\tau$ and $\Lambda_{QCD}$, see \cite{Maris:1999nt}. While the UV-part of this interaction 
is important for systematic reasons, it is known to change the resulting spectrum only slightly as compared to Model I. We therefore in
addition employ a different strategy for variation of parameters: here we keep the interaction scale $\omega_{II} = 0.4$ GeV constant, and
decrease the interaction strength starting from $D_{II} = 0.93$ GeV$^2$, which is the physical point of the model taken from \cite{Maris:1999nt}, down to 
$D_{II} = 0.03$ mGeV$^2$. 
{All calculations have been performed with a current 
mass $m[\mu = 19\, \mathrm{GeV}]=3.7$ MeV.}

For both model I and model II the quark dressing function need only to be solved iteratively for positive real momenta squared $p^2\ge 0$. 
Once they are accurately known there, the values of the dressing functions can be determined for complex values of $p^2$ without further 
need for iterative methods by simply calculating the right-hand-side of the respective integral equation. Note, however, that this is
accidental. As has been noted already in \cite{Maris:1995ns}, this procedure suffers from integrating over cuts produced in the angular 
integral of the quark self-energy. It so happens that these cuts are so small for the Model I and Model II that the corresponding errors
are negligible. This is generically not the case and also does not work for Model III.

\subsection{Model III}

Our third model goes beyond truncations with models for the effective running coupling restricted to the quark sector of QCD. 
Instead, we solve the coupled set of DS equations for the quark, gluon and ghost propagator following the truncation of Ref.~\cite{Fischer:2005en}. In general, the ghost and gluon propagators are represented by  
\begin{align}
    D_{\mu\nu}(k) &=  \left( \delta_{\mu \nu} - \frac{k_\mu k_\nu}{k^2} \right) \frac{Z(k^2)}{k^2} \,, \\
    D_G(k) &= - G(k^2)/k^2 \,,
\end{align}
where $Z(k^2)$ and $G(k^2)$ are the dressing functions of the gluon and ghost, respectively. We again use a rainbow-ladder type 
approximation for the full quark-gluon-vertex, but explicitly take into account a structure of the dressing function that is imposed
by the corresponding Slavnov-Taylor identity (see \cite{Fischer:2003rp,Fischer:2005en} for details, see also Ref.~\cite{Pawlowski:2024kxc} for a similar interaction which replaces $G^2(k^2) \to G(k^2)$ in the study of the corresponding spectral DSE)
\begin{align}
    \Gamma^{\nu,a}(q,p) = \frac{ig Z_2}{Z_{1F}} T^a \gamma_\nu G^2(k^2) A(k^2) \, ,
\end{align}
where $k=p-q$ is the gluon momentum. In the notation of Eq.(\ref{eq:alkofer-watson-weigel}) this results in an 
effective running coupling 
\begin{align}
\mathcal{G}(k^2) = \frac{1}{Z_2} g^2 G^2(k^2) Z(k^2) A(k^2)
\end{align}
which is invariant under changes of renormalization similar to models I and II. 

In order to again explore the model at different interaction-strengths, we multiply the coupling by a scale dependent strength-parameter 
$h(k^2,D_{III})$, that preserves the UV physics while providing damping in the IR:
\begin{align} \label{eq:damping}
    g^2 &\to g^2 h(k^2,D_{III}), \quad \quad   D_{III} \in [0,1], \nonumber \\ 
    h(k^2,D_{III}) &= \arctan\left(\frac{k^2}{10}\right)\frac{2}{\pi}(1-D_{III})+D_{III}.
\end{align}
For our results in section \ref{sec:results} we vary $D_{III}$ between 
$D_{III}=1$ and $D_{III}=0.1$.
{For model III all calculations have been performed with a current 
quark mass $m[\mu = 31\, \mathrm{GeV}]=1.9$~MeV.}

In this approach every Yang-Mills dressing function remains purely real. In order to solve the quark DS equation we pursue the same 
approach as in Ref.~\cite{Fischer:2005en}: First solve all propagator DS equations self-consistently on the real semi-positive 
half-axis, i.e., for $p^2 \in \mathbbm{R}$, $p^2\ge 0$. This procedure also fixes all renormalization constants. Then, we solve 
the momentum shifted equation, i.e., we substitute $q \rightarrow p-q$. The Yang-Mills propagators then only depend
on the real new loop momentum q, ($G(q^2), Z(q^2)$), whereas the internal quark propagator carries the momentum difference
$k=p-q$. The resulting equation can then be solved on a momentum grid with parabola shape in the complex momentum plane.

\section{Meson Bound States and Bethe-Salpeter Equations}\label{sec:BSE}

As already stated above for all three models we use a rainbow-ladder type truncation of the quark propagator DS and meson BS 
equations. In this section we briefly review some aspects of the BS equation for mesons which are of particular relevance for 
the study presented here, for more details about the BS equation and its use in hadron physics we refer, e.g., to the review 
\cite{Eichmann:2016yit}.

The Bethe-Salpeter equation for quark-antiquark states can be written as 
\begin{align}\label{eq:BSE}
	[\Gamma^{(\mu)}(p, P)]_{tu} = \int_q K_{tu,rs}(q, p, P)
	[S(q_+) \Gamma^{(\mu)}(q,P) S(q_-)]_{sr}\,,
\end{align}
with generic indices for color, flavor and Dirac structures of the interaction kernel $K(q,p,P)$. The  
BS-amplitude $\Gamma^{(\mu)}(p,P)$ depends on the total momentum $P$ of the meson and the relative momentum 
$p$ of the quark-antiquark pair. The BS equation can be solved numerically as eigenvalue equation of the form
$\lambda(P^2)\, \Gamma =  (KS_+S_-)\cdot\Gamma$ with $S_\pm = S(q_\pm)$. Note that the momenta $q_\pm = q \pm P/2$
appearing as arguments of the quark propagators are complex, since in the rest frame of the meson we have $P=(0,0,0,iM)$ 
with meson mass $-P^2=M^2$. Evaluating $q_\pm^2$ one finds that the region of complex momenta needed to solve the 
BS-equation is bounded by a parabola with the real line as symmetry axis and with the apex at $-M^2/4$, $M$ denoting 
the respective meson's bound state mass. In practice, one solves the DS-equation in the complex momentum plane once 
on a parabola with maximal possible $\mathcal{M}$ such that poles in the quark propagator are not encountered inside
the parabola. In turn this means that using this quark propagator, the BS-equation can be explored for all mesons with 
masses $M \le \mathcal{M}$ without additional extrapolations in the eigenvalue. This is indeed all we need for this work.   

Depending on the quantum numbers of the meson in question, the amplitude $\Gamma^{(\mu)}$ carries different Dirac 
structures. For pseudoscalar (P) and scalar (S) mesons with quantum numbers $J^{PC}=0^{-+}$ and $J^{PC}=0^{++}$ 
in vacuum the Dirac tensor decomposition of the BS-amplitudes is given by
\begin{align}
	\Gamma_{\textup{P}}(p,P)=\;\gamma_{5}&\left\{E_{\textup{P}}(p,P)-i\slashed{P}F_{\textup{P}}(p,P)\right. \nonumber \\
	&\;-\left.i\slashed{p}\,P\cdotp p\,G_{\textup{P}}(p,P)+\left[\slashed{P},\slashed{p}\right]H_{\textup{P}}(p,P)\right\}, \label{eq:BSA_vacuum_pseudoscalar}\\
	\Gamma_{\textup{S}}(p,P)=\,\;\openone&\left\{E_{\textup{S}}(p,P)-i\slashed{P}\,P\cdotp p\,F_{\textup{S}}(p,P)\right. \nonumber \\
	&\;-\left.i\slashed{p}\,G_{\textup{S}}(p,P)+\left[\slashed{P},\slashed{p}\right]H_{\textup{S}}(p,P)\right\}.
	\label{eq:BSA_vacuum_scalar}
\end{align}
The corresponding flavor and color parts of the amplitudes can be found in Ref.~\cite{Eichmann:2016yit} together with
details on the underlying construction principles as well as explicit expressions for parity and charge conjugation operations. 
In general, the $\gamma_{5}$-factor of the pseudoscalar mesons accounts for parity, whereas extra factors $P \cdotp p$ at 
some places ensure that all BSA components $E_X$, $F_X$, $G_X$, and $H_X$ with $X \in\{\textup{P},\textup{S}\}$
transform similarly under charge conjugation. Furthermore, the signs are chosen such that all BSA components are positive. 

For the vector (V) and axial-vector (A) meson with quantum numbers $J^{PC}=1^{--}$ and $J^{PC}=1^{+-}$ as well as 
$J^{PC}=1^{++}$ in vacuum we work with the tensor decomposition
\begin{align}
	\label{eq:V_AV_BSA_vacuum}
	\Gamma_\textup{V}^\mu(p,P)=&\;i\gamma_\top^\mu F_{1\textup{V}}(p,P)+\gamma_\top^\mu\slashed{P} F_{2\textup{V}}(p,P) \nonumber \\
	&\;+(p_\top^\mu\openone_\textup{D}-\gamma_\top^\mu\slashed{p})\,P\cdotp p\, F_{3\textup{V}}(p,P) \nonumber \\
	&\;+(i\gamma_\top^\mu\commutator{\slashed{P}}{\slashed{p}}+2ip_\top^\mu\slashed{P}) F_{4\textup{V}}(p,P) \nonumber \\
	&\;+p_\top^\mu\openone_\textup{D} F_{5\textup{V}}(p,P)+ip_\top^\mu\slashed{P}\,P\cdotp p\, F_{6\textup{V}}(p,P) \nonumber \\
	&\;-ip_\top^\mu\slashed{p} F_{7\textup{V}}(p,P)+p_\top^\mu\commutator{\slashed{P}}{\slashed{p}} F_{8\textup{V}}(p,P), \nonumber \\[0.5em]
	\Gamma_\textup{A}^\mu(p,P)=&\;\gamma_5\left\{i\gamma_\top^\mu F_{1\textup{A}}(p,P)+\gamma_\top^\mu\slashed{P} \,P\cdotp pF_{2\textup{A}}(p,P) \right. \nonumber \\
	&\;+(p_\top^\mu\openone_\textup{D}-\gamma_\top^\mu\slashed{p})\, F_{3\textup{A}}(p,P) \nonumber \\
	&\;+(i\gamma_\top^\mu\commutator{\slashed{P}}{\slashed{p}}+2ip_\top^\mu\slashed{P}) F_{4\textup{A}}(p,P) \nonumber \\
	&\;+p_\top^\mu\openone_\textup{D} \,P\cdotp pF_{5\textup{A}}(p,P)+ip_\top^\mu\slashed{P}\,P\cdotp p\, F_{6\textup{A}}(p,P) \nonumber \\
	&\;\left.-ip_\top^\mu\slashed{p} F_{7\textup{A}}(p,P)+p_\top^\mu\commutator{\slashed{P}}{\slashed{p}} \,P\cdotp pF_{8\textup{A}}(p,P)\right\}\,,
\end{align}
which is constructed such that the on-shell (axial-)vector meson is transverse to its total momentum $P$. 
The subscript $\top$ indicates transversality of $w\in\{\gamma,p\}$ w.r.t.~the total momentum, i.e., $w_\top^\mu=T_{\mu\nu}(P)w^\nu$. The two different charge conjugation eigenstates of the axialvector mesons
corresponding to the $a1$ and $b1$ states in nature show up with masses close to each other in the numerical 
solutions of the corresponding BS-equation. They are conveniently distinguished using a Chebychev expansion 
of the $p\cdot P$-dependence of the dressing functions $F_{i\textup{A}}$ and taking into account only contributions 
even or odd in powers of $p\cdot P$. 

Finally, we have to specify the scattering kernel $K$ in the Bethe-Salpeter equation. It cannot be chosen arbitrarily, 
but it has to match the truncation scheme used in the DSE such that the axialvector Ward-Takahashi identity is satisfied
and the (pseudo-)Goldstone nature of the pion preserved. Corresponding details and an explicit derivation of this property 
can be found, e.g., in Ref.~\cite{Eichmann:2016yit}.
For the rainbow-ladder truncation, this is achieved by using a single effective gluon exchange between two bare vertices
\begin{equation}\label{eq:rl-kernel}
	K_{tu,rs}(p,q,P) = -C_f Z_2^2 [\gamma^{\mu}]_{tu} [\gamma^\nu]_{rs} \mathcal{G}\left(k^2\right) D_{free}^{\mu\nu}(k)\,,
\end{equation}
with the same effective running coupling $\mathcal{G}$ as the one in the quark self-energy detailed above for our models I-III. 

\section{Numerical results}\label{sec:results}

In the following we will detail our results for the analytic structure of the quark propagator and the masses of 
(pseudo-)scalar and (axial-)vector mesons as a function of the interaction strength in all three models I-III. We
first discuss the pole structure of the quark propagator, but ask the reader to keep in mind that we will subsequently establish 
a direct connection to the emerging meson masses and mass degeneracies.   

\subsection{Pole structure of the quark propagator}

When solving the respective DS equations we note that in all three models we find a qualitatively similar analytic structure of the quark propagator as a function of interaction strengths. The quark DS equation for models I and II can be solved in the entire complex plane. For model I, we can even closely track the poles by computing the associated pole-counting integrals, and determine the multiplicity of poles. In model III, we find that we can solve the DSEs self-consistently beyond the occurrence of the first pole, as long as the first pole is real. However, our system of DSEs fails to converge if the set of external momenta contains a complex pole.

For very small strength parameters $D_{I,II,III}$ we always find a real-valued pole of the quark propagator at or close to
the origin. The typical momentum scale of the pole is of the order of the current quark mass, $p^2 \sim -m^2$ 
with $m \rightarrow 0$ in the chiral limit. This is, of course, what is expected when the interaction strength is not
large enough to trigger dynamical chiral symmetry breaking. In model I and II, we observe that the next closest pole 
is actually a degenerate pair of real-valued poles. In model I, we can numerically confirm this by using a pole-counting integral which {results for this singularity in a value of} approximately two. In model II this does not work reliably due to the logarithmic UV part in the interaction. However, we see {that} these poles become two nearly-degenerate poles as chiral symmetry is broken and therefore conclude that for the chirally symmetric scenario {also} two degenerate poles are present. We can only conjecture that this behavior persists for model III, but we do not have numerical evidence since we are unable to solve the DSE beyond the first pole.

The first pole remains close to the origin until either $D_{I,II,III}$ is large enough to trigger dynamical chiral 
symmetry breaking. At this point the lowest-lying pole starts to move along the real axis deeper into the timelike 
momentum region. For model I and II, we observe that the degeneracy in the next real pole pair is lifted as chiral 
symmetry gets broken. Eventually, one of the poles moves towards the origin while the other one moves even deeper into 
the timelike region. 

For models I and II we clearly observe that two lowest-lying real poles will eventually \textit{``collide"} on the real axis, and form a complex conjugate pair of poles. In model I, we are able to see this effect repeated at a larger strength parameters. For model III we can only conjecture this since we only can solve the complex DSEs up to the first pole. However, the fact, {that} we see a pair of complex conjugate poles emerge, suggests that model III behaves similarly {in this respect to} models I and II.
NB: A similar behavior of real poles colliding and forming pairs of complex conjugated poles has been observed when tuning the quark mass in Ref.~\cite{Windisch:2016iud}.

\begin{figure}[t]
    \centering
    \includegraphics[height=0.45\linewidth]{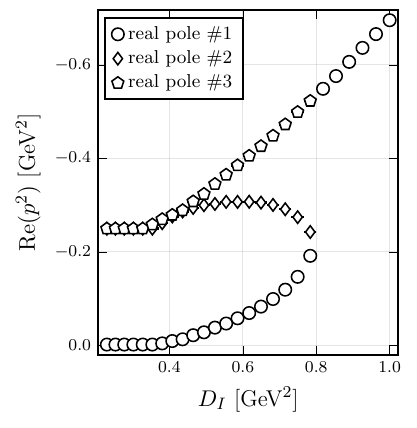}
    \includegraphics[height=0.45\linewidth]{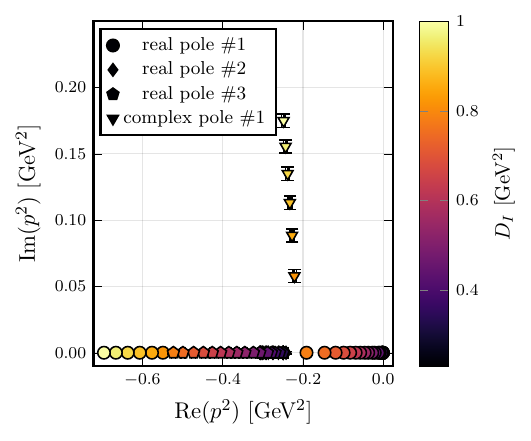}
    \caption{(left) The location of the poles of the function $\sigma(p^2)$ \eqref{sigmaA} for model I in the complex $p^2$-plane as a function of $D_I$, cf. table \ref{tab:param_W}. (right) Position of the poles of $\sigma (p^2)$ along the real axis as a function of $D_I$. When the two low-lying poles cross they generate a pair of complex conjugate poles, cf.\ the left panel. Note, that we also observe a crossing of two real valued poles that does not lead to the formation of a complex pole pair.}
    \label{fig:pole_positions}
\end{figure}
\begin{figure}[t]
    \centering
    \includegraphics[height=0.55\linewidth]{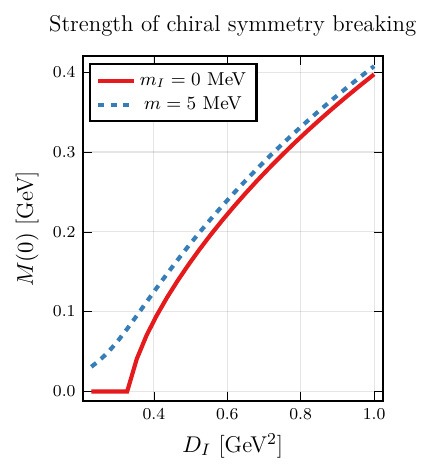}
    \includegraphics[height=0.49\linewidth]{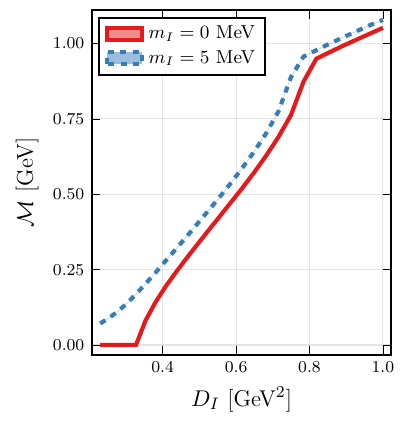}
    \caption{(left) Displayed is the infrared value of the quark mass function $M(0)$, as a proxy for the strength of dynamical chiral symmetry breaking, for model I as function of $D_I$ in the chiral limit and for a small quark current mass. (right) The parameter $\mathcal{M}$ determined by the largest parabola with apex $-\mathcal{M}^2/4$ that is free of poles for the given coupling strength $D_I$. In the presence of non-vanishing current quark masses, the second-order transition, visible in both variables, is smoothened into a crossover behavior.}
    \label{fig:chi_symm_breaking}
\end{figure}
\begin{figure}[t]
	\centering
	\includegraphics[width=0.49\linewidth]{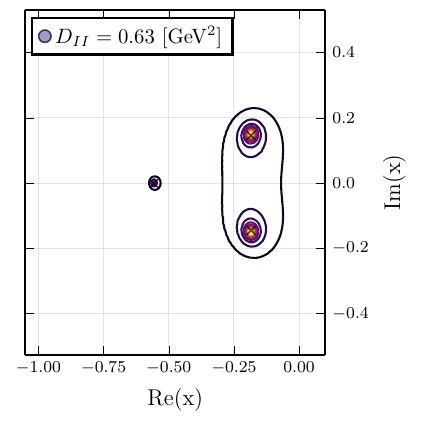}
	\includegraphics[width=0.49\linewidth]{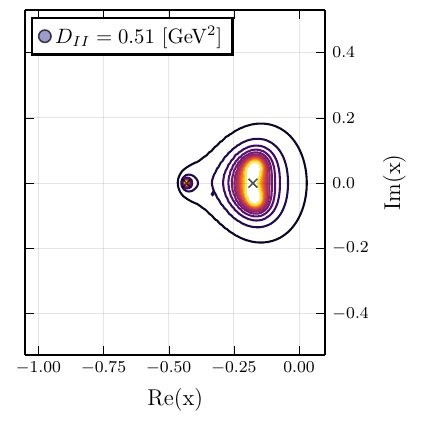} 
	\includegraphics[width=0.49\linewidth]{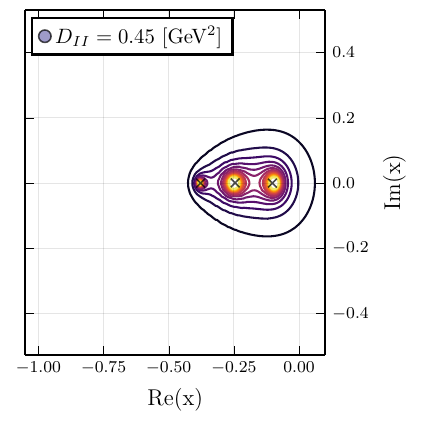} 
	\includegraphics[width=0.49\linewidth]{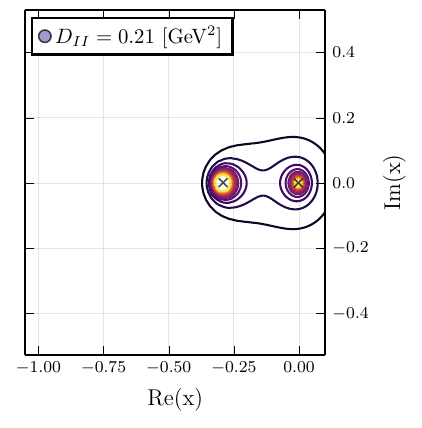} 
	\caption{Contour plot of $\sigma(p^2)$ for model II showing the pole structure emerging as a function of the strength parameter $D_{II}$ for fixed $\omega_{II} = 0.4$ GeV. We observe a pattern similar to the one of model I. We indicate the pole locations with crosses.}
	\label{fig:maris_tandy_strength}
\end{figure}

For illustrating the above described pattern of dependencies of pole locations w.r.t. the interaction strength 
we focus first on the simplest interaction model, i.e., model I {in the chiral limit}. In Fig.~\ref{fig:pole_positions} we show the 
positions of the three lowest-lying quark propagator's poles for complex values of $p^2$ (= closest to the origin) 
for massless quarks, color coded according to the corresponding value of the strength parameter 
{$D_I$}. The pole locations are determined by the zeros of the denominators of the function 
$\sigma_A(p^2)$ \eqref{sigmaA} or, equivalently, $\sigma_B(p^2)$ \eqref{sigmaB}. Poles in the dressing 
functions $A(p^2)$ and/or $B(p^2)$ are irrelevant, since they appear much further out in the time-like 
momentum plane. At $D_I = 0.23$ GeV$^2$ we start with only two pole locations at $p^2=0$ and $p^2=-$ 0.249 GeV$^2$, 
since the latter location accommodates the pair of degenerate poles already discussed above. For 
{$D_I/\omega_I^2 \ge 1.92$} (corresponding roughly to $D_I \approx 0.332$ GeV$^2$) dynamical chiral symmetry breaking occurs leading to the split-up of the degenerate 
poles as visible in the right diagram of Fig.~\ref{fig:pole_positions}. At this point the quark mass function
$M(p^2) = B(p^2)/A(p^2)$ starts to be non-vanishing also in the chiral limit. In the left panel of 
Fig.~\ref{fig:chi_symm_breaking} we display the corresponding value of $M(0)$ as function of the strength 
parameter $D_I$. In the chiral limit, the transition is a second order (quantum) phase transition, whereas at finite quark current quark masses it becomes a cross-over as also displayed in this figure. 

As already discussed above, the poles begin to move with increasing interaction strength but stay real 
in the interval {$0.332 \mathrm{GeV}^2\le D_I \le 0.801 \mathrm{GeV}^2 $}. As can be seen from Fig.~\ref{fig:pole_positions}, two poles
merge at the upper end of this interval and they split up into a pair of complex-conjugated poles
when further increasing the interaction strength. The appearance of complex-conjugated poles does not have 
any visible effect on the quark propagator dressing functions for Euclidean momenta. As can be seen in the 
left panel of Fig.~\ref{fig:chi_symm_breaking} $M(0)$ is a smooth function of the interaction strength 
$D_I$ in the respective region. 

In the right panel of Fig.~\ref{fig:chi_symm_breaking} we show the parameter $\mathcal{M}$ 
defined via the largest parabola with the real axis as symmetry axis and an apex at $-\mathcal{M}^2/4$ that remains 
free of poles, cf. our discussion below Eq.~(\ref{eq:BSE}). Of course, by construction $\mathcal{M}$ shows a cusp 
when the complex-conjugated pole pair appears, since then the real part of the pole location hardly changes whereas
mainly the imaginary part grows leading only to slight larger possible parabolas with growing interaction strength.

For further illustration we also show contour plots of $\sigma(p^2)$ in the complex momentum plane for four selected values of the interaction strength $D_{II}$ of model II in Fig.~\ref{fig:maris_tandy_strength}. {The emerging pole structure is clearly visible. We indicate the locations of the poles, determined from the local (numerical) maxima of $\sigma(p^2)$, with crosses.} In the lower right panel, 
the interaction strength is subcritical, and therefore the quark propagator possesses a pole at the origin and a 
double pole at finite $p^2$. Increasing the interaction strength one has, as in model I, first three real poles 
(lower left panel) and then one pair of complex-conjugated poles and a real pole (upper panels).  
The main difference is that also $\mathcal{G}_{\rm UV}(k^2)$ \eqref{eq:maris-tandyUV} provides some interaction 
strength, and therefore dynamical chiral symmetry breaking and pole merging happen at respective smaller values 
of the IR interaction strength. 

\subsection{Meson mass degeneracies}
\begin{figure*}[t]
    \centering
    \includegraphics[width=0.30\linewidth]{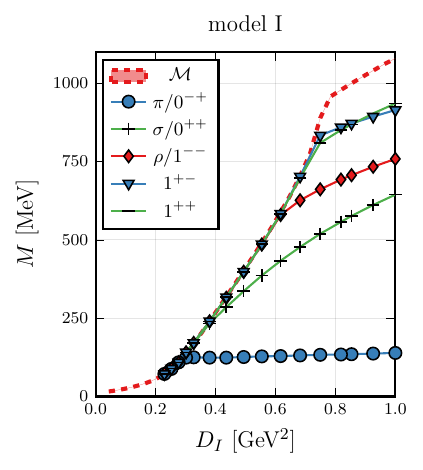}~
    \includegraphics[width=0.30\linewidth]{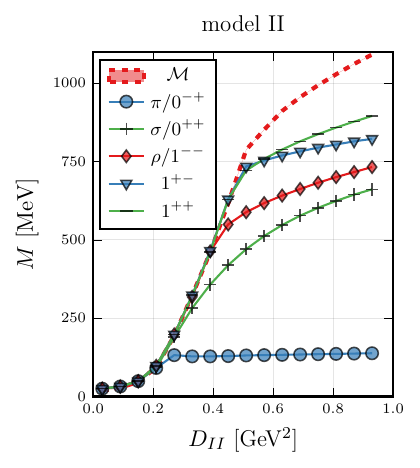}~
    \includegraphics[width=0.30\linewidth]{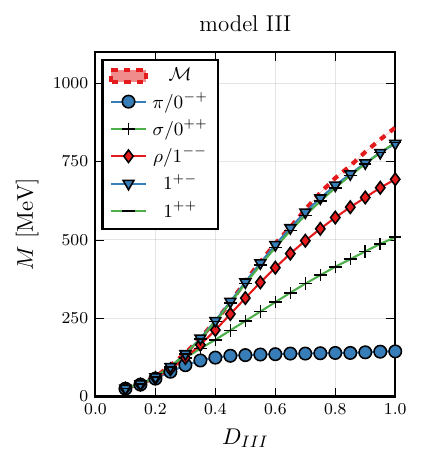}
    \caption{Masses of mesons with quantum numbers $J^{PC} = 0^{-+}, 0^{++}, 1^{--}, 1^{++}, 1^{+-}$ as functions of the strength parameters $D_{I,II,III}$ of models I-III. In all cases we obtain the same pattern of degeneracy at small interaction strengths. Shown is furthermore the parameter $\mathcal{M}$ characterizing the largest parabolic region symmetric to real $p^2$ with apex $-\mathcal{M}/4$. }
    \label{fig:meson_masses}
\end{figure*}
\begin{figure}[t]
	\centering
	\includegraphics[width=0.49\linewidth]{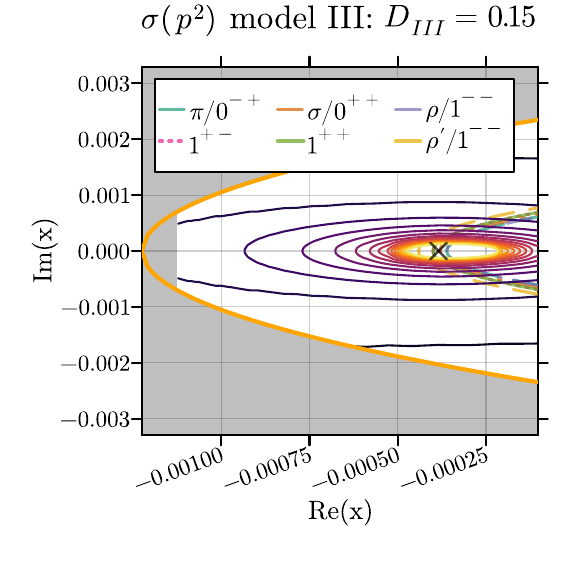}
	\includegraphics[width=0.49\linewidth]{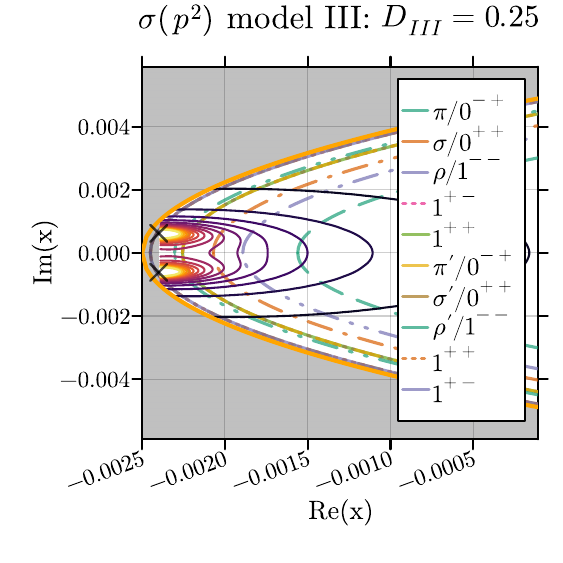}
	\includegraphics[width=0.49\linewidth]{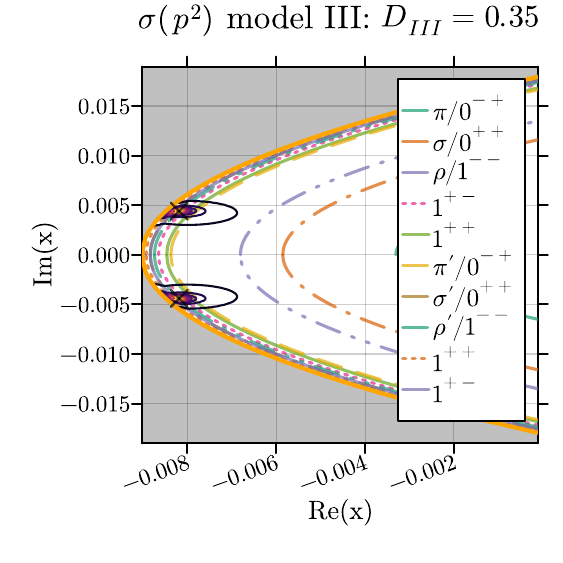}
	\includegraphics[width=0.49\linewidth]{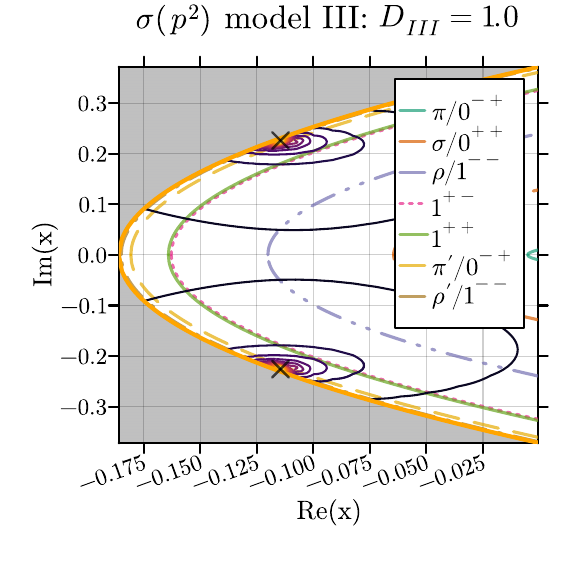}
	\caption{Dressing function $\sigma(p)$ for model III based on the ghost and gluon propagators on the real axis for various strength parameters $D_{III}$. The plot shows the largest parabola for which we could solve the complex quark DSEs self-consistently, as well as the parabolas associated with the meson masses as determined through their respective BSEs. We indicate the pole locations with crosses.}
	\label{fig:quark_poles_model_iii}
\end{figure}
When solving the meson BS equations for varying interaction strengths we noticed a very interesting general pattern, 
see Fig.~\ref{fig:meson_masses}. For small values of the interaction strength we observed degeneracies in the low-energy 
meson spectrum. Furthermore, these degeneracies appear to be linked to the position of the quark propagator's closest 
pole to the origin. The observed degeneracies are hereby qualitatively similar in all three models and differ only 
somewhat quantitatively.

For a detailed discussion let us start at the physical point, i.e., at maximal interaction strength. All three diagrams 
in Fig.~\ref{fig:meson_masses} show the masses of mesons with quantum numbers $J^{PC} = 0^{-+}, 0^{++}, 1^{--}, 1^{++}, 1^{+-}$
in the respective models and with current quark masses adapted to reproduce the experimental value of the mass of the 
pion. As is typical for rainbow-ladder type approximations, the vector meson masses are not too far away from their 
experimental value (although there is no width), whereas the masses of the scalar meson and the axialvectors are far 
off. This is well-known and discussed in detail e.g. in Ref.~\cite{Eichmann:2016yit}. Beyond rainbow-ladder truncations
(partly) remedy this situation. However, these require orders of magnitude more numerical effort and are therefore 
currently not feasible for the purpose of this exploratory work. We believe these are also not necessary, since we 
are mainly interested in emerging general properties such as degeneracy patters and their relations to analytical structure.
This purpose is served, to our mind, by the available models. 

Lowering the interaction strength we observe a similar pattern in all three models. First the two axialvector states 
become degenerate (if not already from the start), then the vector mesons joins, then the scalar meson for even lower
interaction strength and finally also the pion becomes mass degenerate with the others once chiral symmetry is restored. 
While the latter phenomenon is naturally explained by chiral symmetry restoration, the approximate degeneracy of the 
other chiral partners, the vector meson (the $\rho$) and the $1^{+-}$ (the $a_1$), already happens at much larger 
values of the interaction strength. Furthermore, the extension of the symmetry to the scalar state and the other 
axialvector is not explained by the `conventional' notion of chiral partners. Such extended degenerate multiplets,
however, have only been postulated in chiral spin symmetry scenarios \cite{Glozman:2022lda,Glozman:2022zpy} in the context
of finite temperature QCD.

It is also very interesting to compare this emerging pattern of degeneracies with the size of the parabola characterised
by $\mathcal{M}$, also shown in the plots. Most directly for model I, but also visible for models II and III we observe that 
the degeneracy pattern seems to be driven by the (real) pole locations of the quark propagator. Once the interaction strength
is small enough that the pair of complex conjugate poles hit the real axis (indicated by the kink in $\mathcal{M}$, cf. 
Fig.~\ref{fig:chi_symm_breaking}), the movement of the location of the pole towards the origin forces the mass curves
to compress and become degenerate. A possible explanation for this behaviour is given in the summary. In any case, to 
our knowledge this link between analytic structure and mass degeneracies has not been observed before and constitutes 
a potential new aspect in the discussion of chiral spin symmetry scenarios \cite{Glozman:2022lda,Glozman:2022zpy}. 

Finally, in model III we also studied the fate of radially excited states. In Figs.~\ref{fig:quark_poles_model_iii} we 
show contour plots of $\sigma(p^2)$ with the emerging pole structure and the parabolas corresponding to the meson masses as determined through the BSEs. The 
masses for the ground states are the same as displayed in Fig.~\ref{fig:meson_masses}, but we also show the corresponding
parabolas for the first radial excitations. For small values of the strength parameter $D_{III}$ the respective first 
excited states are right above the ground states and (at least partly depending on the value of $D_{III}$) degenerate
within numerical accuracy. The same presentation for more values of $D_{III}$ are 
collected in appendix~\ref{sec:appendix_more_plots}.

\section{Conclusion and Outlook}\label{sec:conclusion}

Herein we studied the location of the low-lying quark propagator's poles
and the low-lying meson spectrum in three interaction models 
within the DS-BS approach. These three models are of different
sophistication and formulated such that the interaction 
strength in the sub-GeV region can be controlled by a single parameter. 

A solution of the DS and BS equations in rainbow-ladder truncation reveals several degeneracy patterns in the meson spectrum at small interaction strengths. For an interaction strength which is too small to trigger dynamical chiral symmetry breaking the observed degeneracies are no surprise but the expected ones. For interaction strengths right above the critical one, however, the observed degeneracies are not at all anticipated. We note that with increasing complexity of the model the parameter regions where these degeneracies are observed gets smaller, however, these degeneracies are observed for all three models. More importantly, the degenerated meson mass is closely related to the location of the lowest-lying pole 
of the quark propagator. 

The latter observation can be traced back to the mathematical
properties of the BS equation in ladder approximation. The respective BS kernel contains two quark propagators, and therefore if the integration region comes close to a pole of the
quark propagator this close-by subregion provides a large if not the dominating contribution. {(NB: This mechanism of dominating sub-regions close to poles is at work independently of whether the poles are located on the real axis or occur as pairs of complex-conjugated poles.)} Therefore, independent of the quantum number of the mesons and thus the details of the BS kernel similarly large contributions will be picked up in the integrand.

Due to this, the degeneracies in the meson spectrum are caused
by the analytic structure of the BS kernel irrespective of any symmetry considerations. 

We want to note that ``explanations'' based on the analytic structure are quite common. E.g., the so-called Silver blaze
property \cite{Cohen:2003kd,Gunkel:2019xnh,Gunkel:2019enr,Cohen:2026pzh}, which states that in quantum systems at non-vanishing density observables can only change with density if the chemical potential exceeds the mass of the lightest excitation, can be traced back to the fact that values of certain integrals only vary if the analytic properties of the integrands are modified, cf.\ Sect.~7 of \cite{Fetter71}.

For QCD at vanishing temperature and density, the strength of the quark-gluon interaction, e.g., in the Landau gauge, is of a given certain momentum-dependence and strength. 
{The general formalism to study this interaction strength and the resulting consequences
in the DSE framework at finite temperature and density is available and summarized in
Ref.~\cite{Fischer:2018sdj}. High quality results for various observables have been obtained 
and systematically compared with lattice QCD, see Ref.~\cite{Fischer:2026uni} for a recent 
overview. Indeed, increasing the temperature even slightly beyond the chiral crossover decreases 
the interaction strength considerably. This already shows up in the vector coupling studied in 
this work, but is even more relevant for those tensor structures of the quark-gluon vertex that are identically zero in a chirally symmetric phase (as, e.g., a scalar coupling). Thus one may expect that the mechanism
identified in this work may carry over to the chirally restored high-temperature regime of QCD.
However, quantitatively reliable results in this matter cannot be obtained from simple models.
In high quality truncations, which explicitly include the back-reaction of quarks onto the 
Yang-Mills sector, meson masses and decay constants for various quantum numbers have been 
calculated from BSEs at finite chemical potential and discussed in
Refs.~\cite{Gunkel:2019xnh,Gunkel:2020wcl}. The generalization of this framework to also 
include finite temperature is subject to future work. 
}

{In lattice calculations one already observes degeneracies between, e.g., mesonic vector 
and axialvector correlators at temperatures above the chiral cross-over, see, e.g., 
\cite{Rohrhofer:2019qwq}.} In view of the results 
presented here, a re-examination of these observed degeneracies with respect to their relation 
to the analytic properties of the quark propagator and bound state kernels at these 
temperatures seems worthwhile. Such an investigation may well add to a further understanding 
of the physics and symmetries of strongly interacting matter in the temperature regime around 
and slightly above the chiral crossover. 

\section*{Acknowledgments}
{We are} grateful to Owe Philipsen and Lorenz von Smekal for fruitful discussions.

F.Z. is supported by the STFC Consolidated Grant No.~ST/X000648/1 and acknowledges support from the Advanced ERC grant ERC-2023-ADG-Project EFT-XYZ.

CF acknowledges support by the Deut\-sche Forschungsgemeinschaft (DFG, German Research Foundation) through
the Collaborative Research Center TransRegio CRC-TR 211 ``Strong-interaction
matter under extreme conditions''.

\bigskip

\bigskip

\newpage

{\bf Open Access Statement}---For the purpose of open access, the authors have applied a Creative Commons Attribution (CC BY) licence  to any Author Accepted Manuscript version arising.

{\bf Research Data Access Statement}---The code and data used to generate the plots and tables in this manuscript can be downloaded from Ref.~\cite{coderelease}.

\bigskip

\bigskip

\appendix

\section{More plots and tables of the pole structure}\label{sec:appendix_more_plots}
\begin{figure*}
    \centering
    \includegraphics[width=0.24\linewidth]{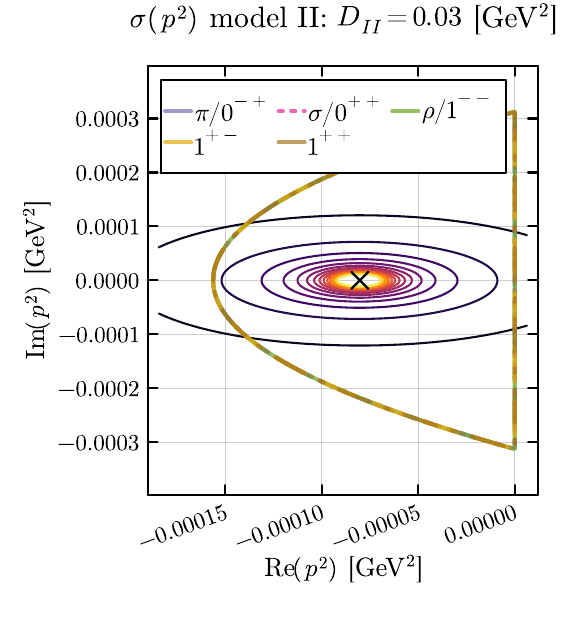}
    \includegraphics[width=0.24\linewidth]{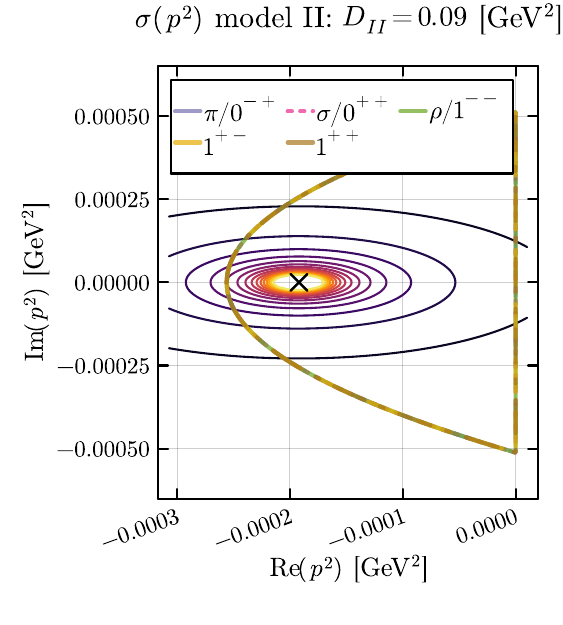}
    \includegraphics[width=0.24\linewidth]{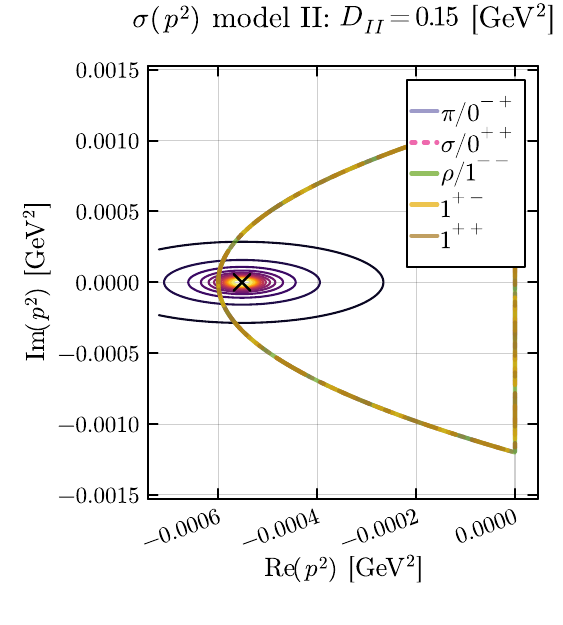}
    \includegraphics[width=0.24\linewidth]{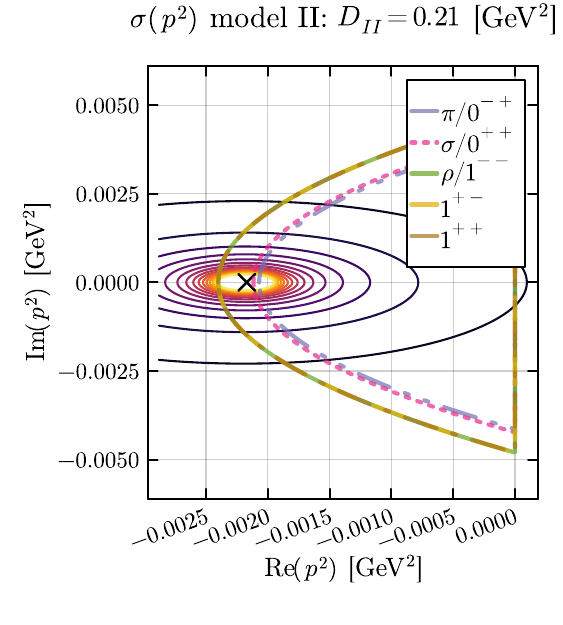}
    \includegraphics[width=0.24\linewidth]{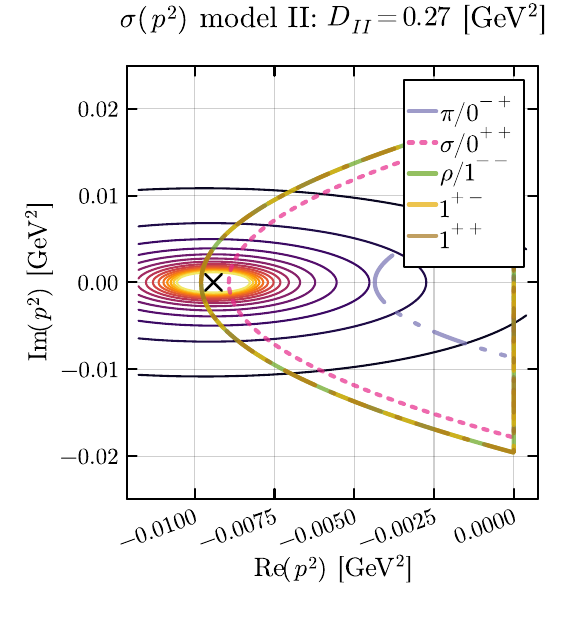}
    \includegraphics[width=0.24\linewidth]{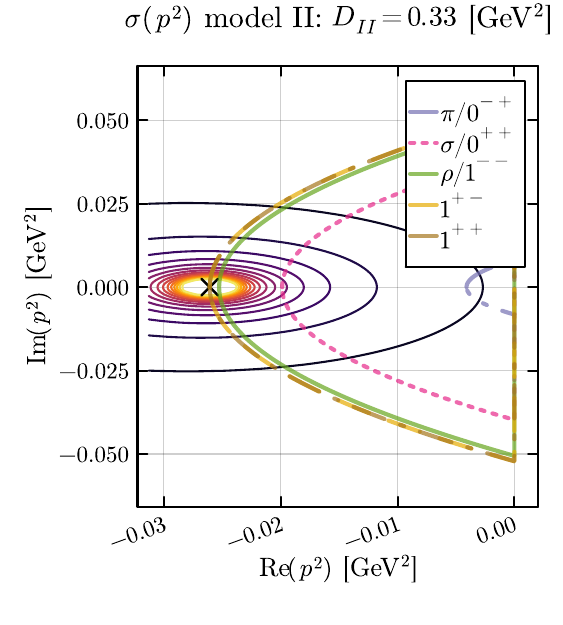}
    \includegraphics[width=0.24\linewidth]{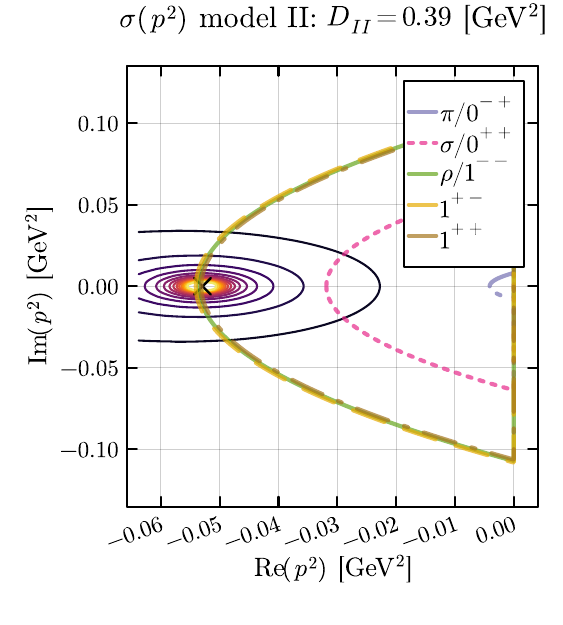}
    \includegraphics[width=0.24\linewidth]{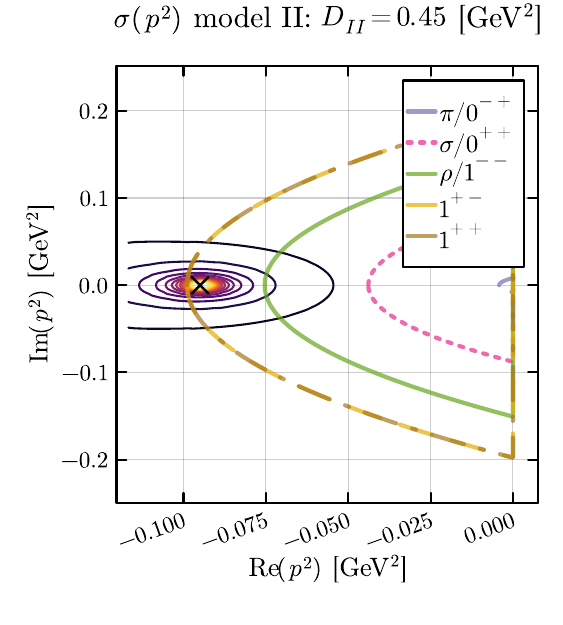}
    \includegraphics[width=0.24\linewidth]{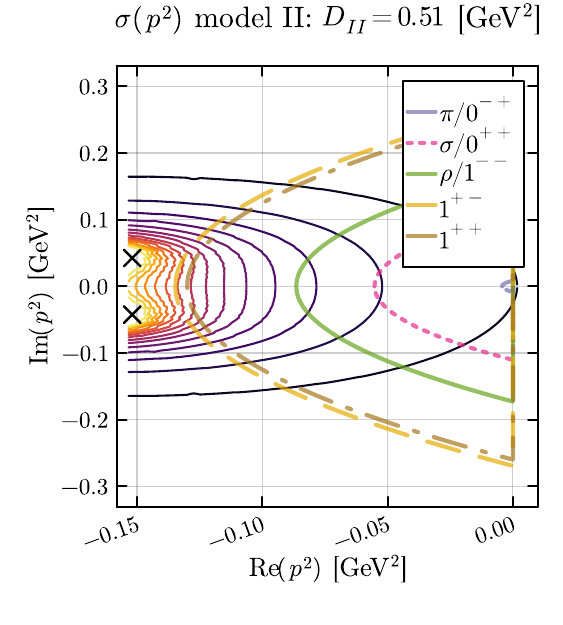}
    \includegraphics[width=0.24\linewidth]{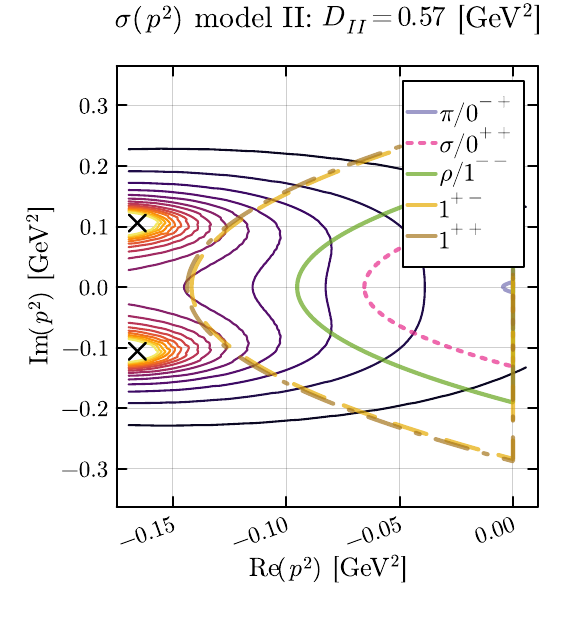}
    \includegraphics[width=0.24\linewidth]{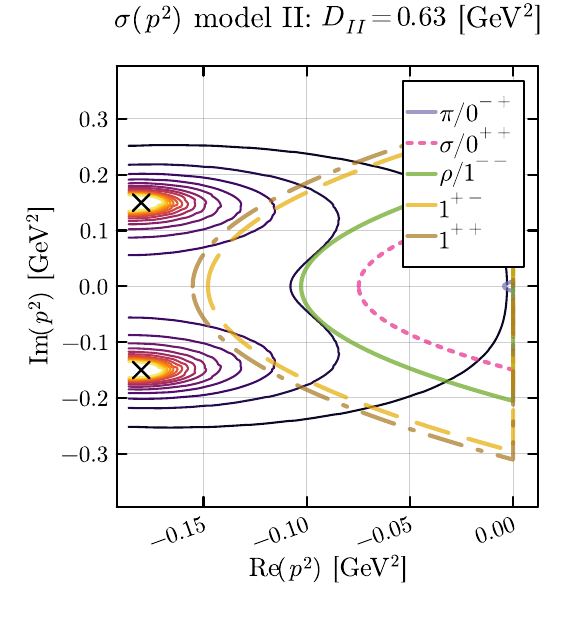}
    \includegraphics[width=0.24\linewidth]{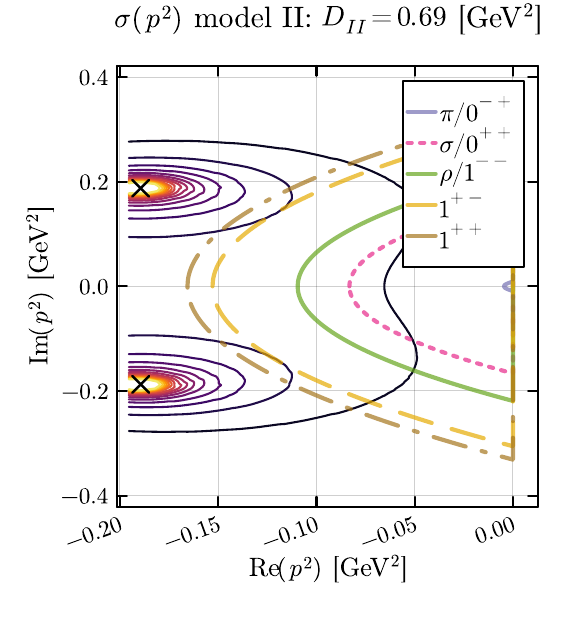}
    \includegraphics[width=0.24\linewidth]{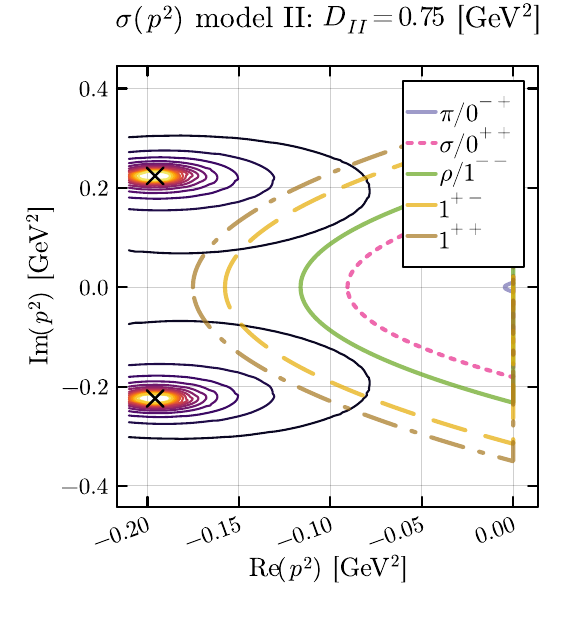}
    \includegraphics[width=0.24\linewidth]{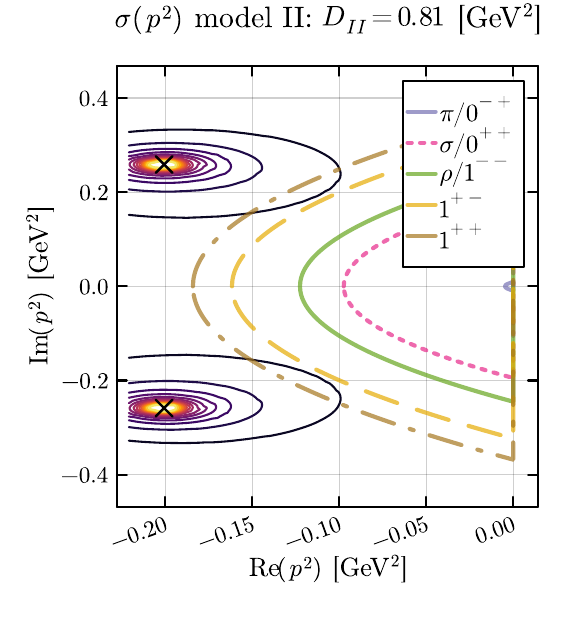}
    \includegraphics[width=0.24\linewidth]{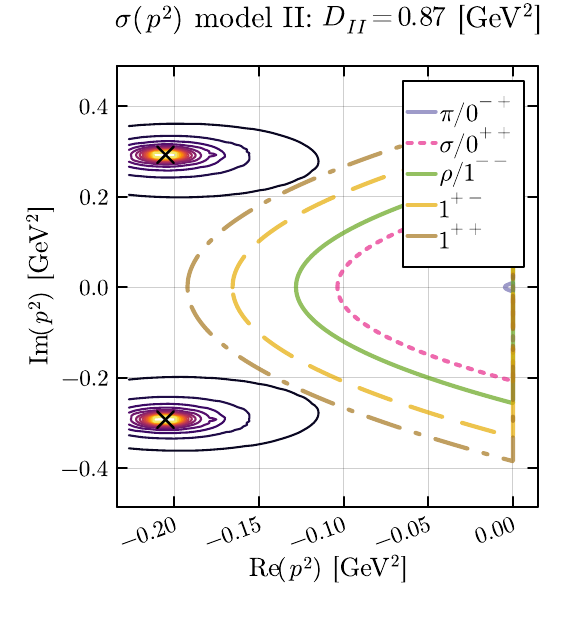}
    \includegraphics[width=0.24\linewidth]{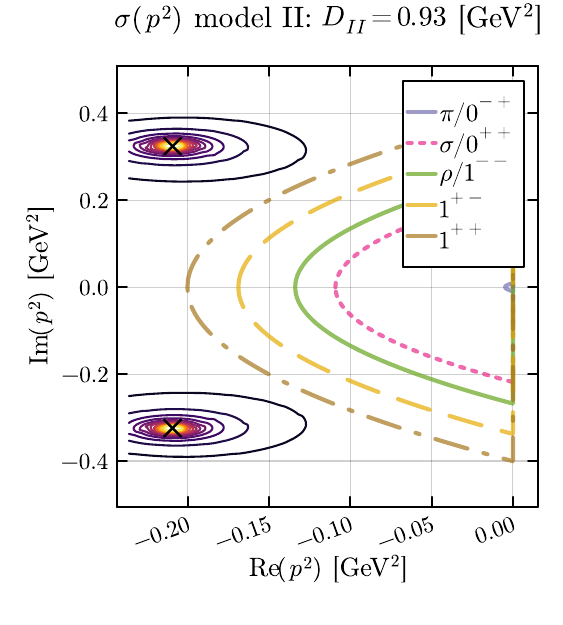}
    \caption{Dressing function $\sigma(p)$ for model II based on the Maris-Tandy interaction for a fermion mass of $3.7$ MeV. The model parameter $\omega_{II}=0.4$ GeV is fixed while the parameter $D_{II}$ is varied. We show the parabolas corresponding to the meson masses as determined through their respective BSEs. We indicate the pole locations with crosses.}
    \label{fig:all_plot_model_II}
\end{figure*}

\begin{figure*}
    \centering
    \includegraphics[width=0.23\linewidth]{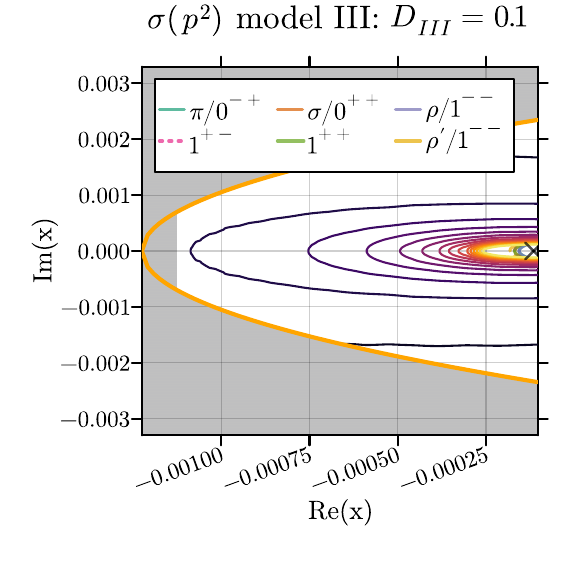}
    \includegraphics[width=0.23\linewidth]{plots/QCD/data_f0.15M0.07_contour.pdf}
    \includegraphics[width=0.23\linewidth]{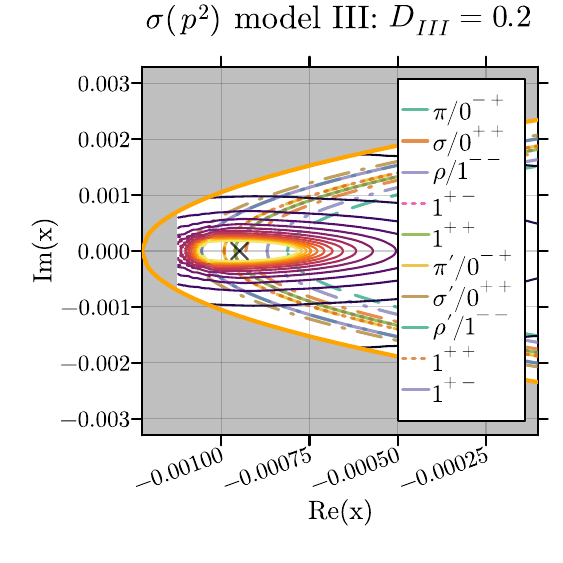}
    \includegraphics[width=0.23\linewidth]{plots/QCD/data_f0.25M0.10_contour.pdf}
    \includegraphics[width=0.23\linewidth]{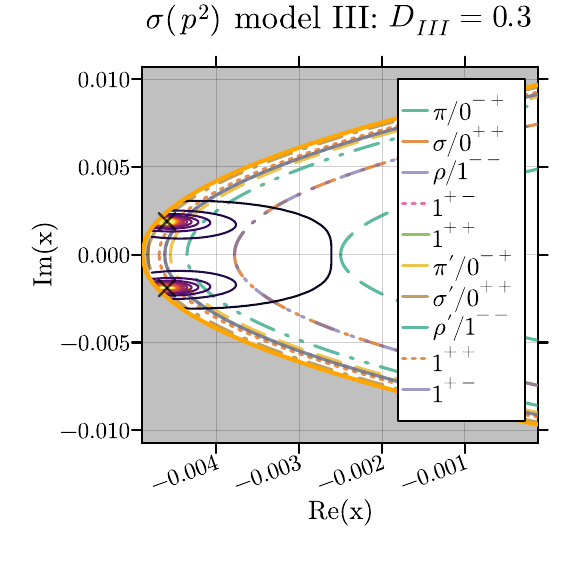}
    \includegraphics[width=0.23\linewidth]{plots/QCD/data_f0.35M0.19_contour.pdf}
    \includegraphics[width=0.23\linewidth]{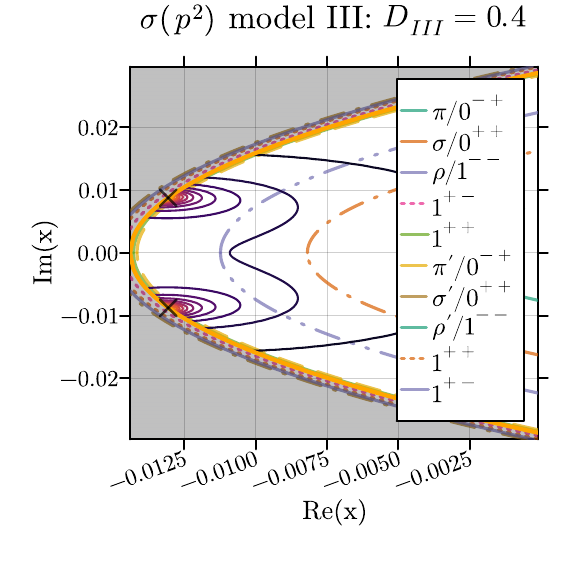}
    \includegraphics[width=0.23\linewidth]{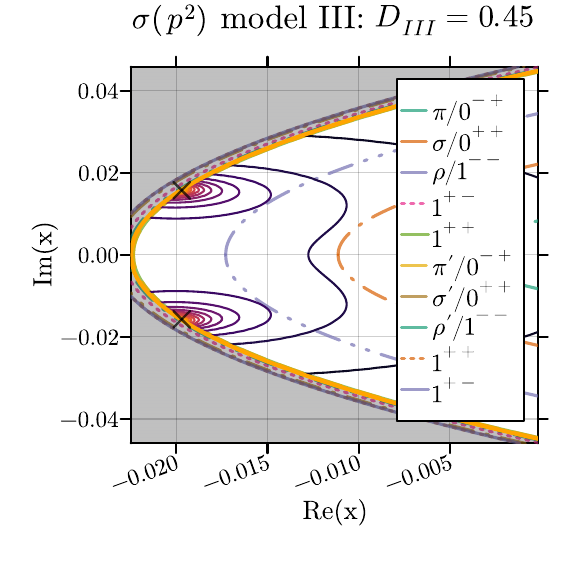}
    \includegraphics[width=0.23\linewidth]{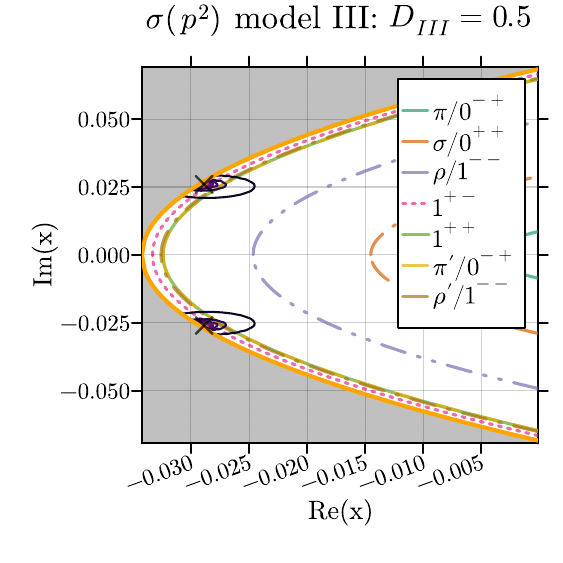}
    \includegraphics[width=0.23\linewidth]{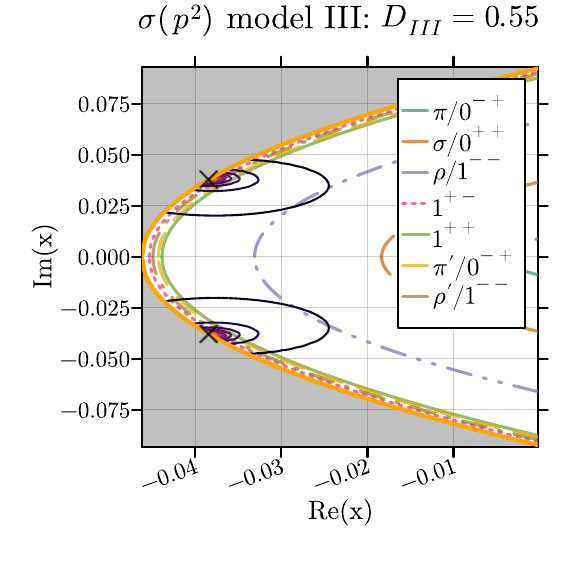}
    \includegraphics[width=0.23\linewidth]{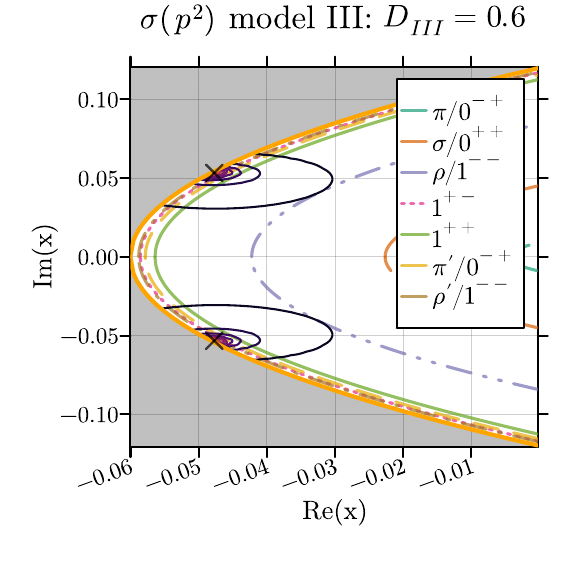}
    \includegraphics[width=0.23\linewidth]{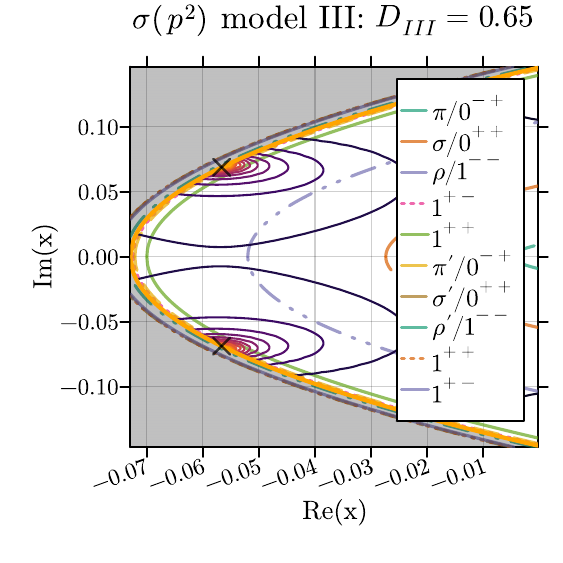}
    \includegraphics[width=0.23\linewidth]{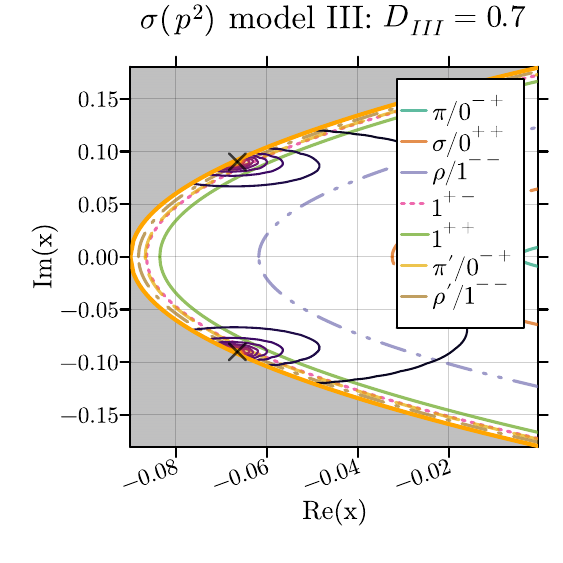}
    \includegraphics[width=0.23\linewidth]{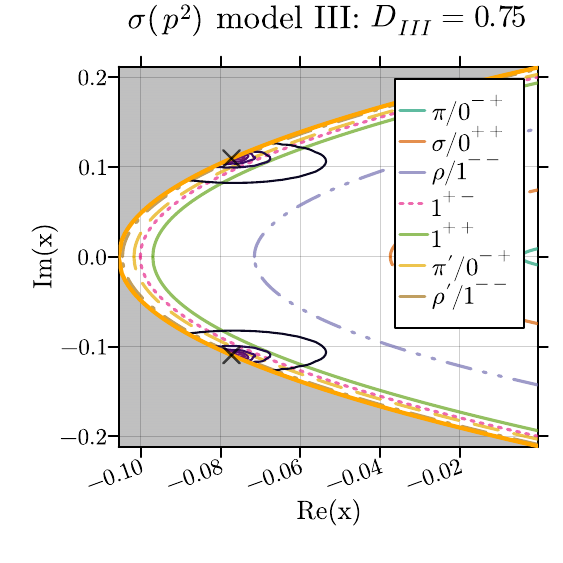}
    \includegraphics[width=0.23\linewidth]{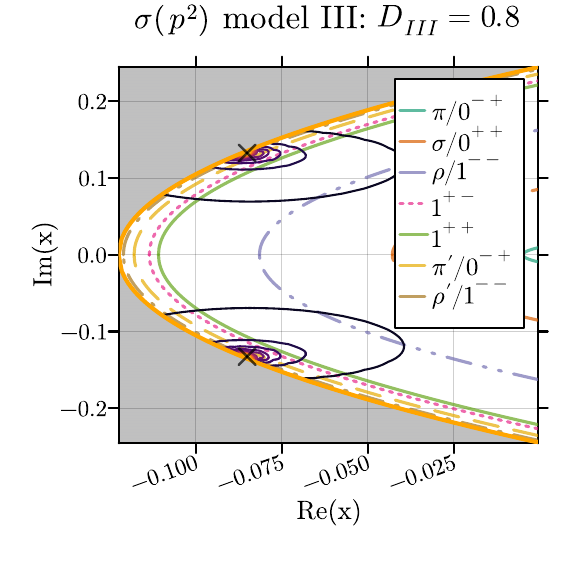}
    \includegraphics[width=0.23\linewidth]{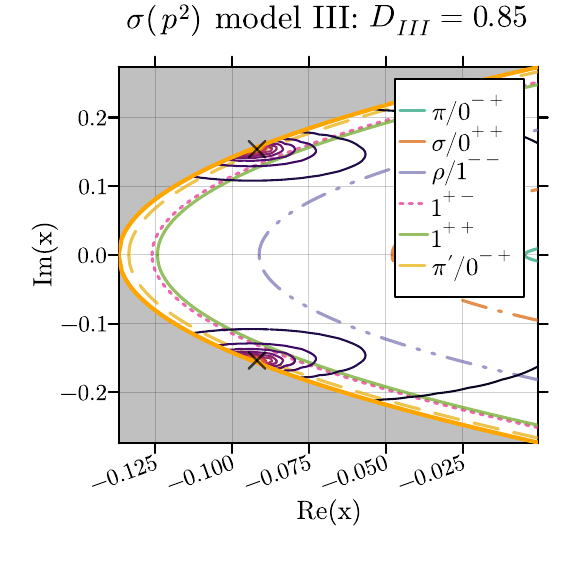}
    \includegraphics[width=0.23\linewidth]{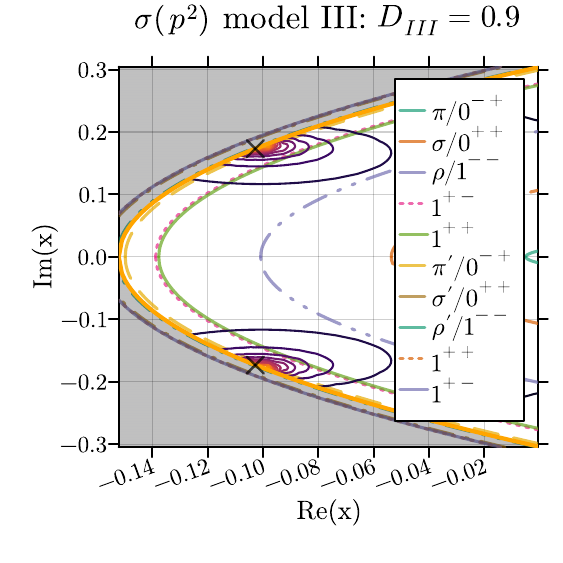}
    \includegraphics[width=0.23\linewidth]{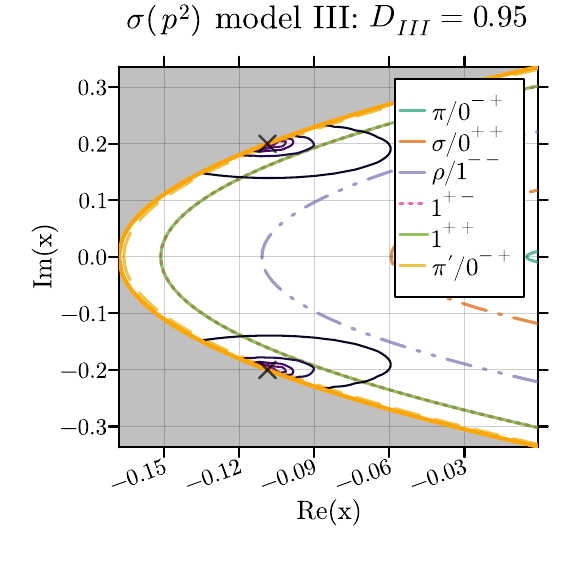}
     \includegraphics[width=0.23\linewidth]{plots/QCD/data_f1.00M0.8625_contour.pdf}
    \caption{Dressing function $\sigma(p)$ for model III based on the ghost and gluon propagators on the real axis for various strength parameters $D_{III}$. The plot shows the largest parabola for which we could solve the complex quark DSEs self-consistently, as well as the parabolas associated with the meson masses as determined through their respective BSEs. For $D_{III}<0.2$ we can extend the parabolic region in the DSE iteration procedure and retain convergence. We indicate the pole locations with crosses.}
    \label{fig:all_plot_model_III}
\end{figure*}

We report the parameters used for model I in Tab.~\ref{tab:param_W}. We show plots of the pole structure of all parameters investigated in this work. In Fig.~\ref{fig:all_plot_model_II} we show the plots of the pole structure for model II, and in Fig.~\ref{fig:all_plot_model_III} we show plots for model III. We further tabulate all data depicted in Fig.~\ref{fig:meson_masses} in Tabs.~\ref{tab:model_I_and_II} and \ref{tab:model_III}.

\begin{table*}[!th]{
\begin{tabular}{|c||c|c|c|c|c|c|c|c|c|c|c|c|c|c|c|c|c|}\hline\hline
$D_I$ & 1 & 0.926 & 0.854 & 0.818 & 0.749 & 0.682 & 0.617 & 0.555 & 0.494 & 0.436 & 0.380 & 0.327 & 0.302 & 0.277 & 0.253 & 0.230 \\\hline
$\omega_I$ & 0.500 & 0.494 & 0.487 & 0.484 & 0.477 & 0.469 & 0.461 & 0.453 & 0.445 & 0.435 & 0.426 & 0.415 & 0.410 & 0.404 & 0.398 & 0.391 \\\hline
$D_I/\omega_I^2$ & 4 & 3.8 & 3.6 & 3.5 & 3.3 & 3.1 & 2.9 & 2.7 & 2.5 & 2.3 & 2.1 & 1.9 & 1.8 & 1.7 & 1.6 & 1.5 \\\hline\hline
\end{tabular}
\caption{Parameters used in model I. {The units are GeV$^2$ for $D_I$ and GeV for $\omega_I$.}
\label{tab:param_W}}
}
\end{table*}

\begin{table*}[h]
    \centering
    \caption{(left) Tabulated values of the meson masses extracted from the BSEs for model I set to a physical scale {(MeV)} and the corresponding parameter {$\mathcal{M}$}. (right) Tabulated values of the meson masses extracted from the BSEs for model II set to a physical scale {(MeV)} and the corresponding parameter $\mathcal{M}$.}
    {\begin{tabular}[t]{|c|c|c|c|c|c|c|c|}
\hline \hline
$D_{I}$ & $m(0^{-+})$ & $m(0^{++})$ & $m(1^{--})$ & $m(1^{+-})$ & $m(1^{++})$ & $\mathcal{M}$ \\ \hline \hline
0.230 & 72 & 72 & 73 & 72 & 72 & 72 \\ 
0.253 & 88 & 88 & 89 & 89 & 89 & 89 \\ 
0.277 & 109 & 110 & 109 & 109 & 109 & 111 \\ 
0.302 & 124 & 140 & 141 & 141 & 141 & 137 \\ 
0.327 & 125 & 170 & 171 & 171 & 171 & 169 \\ 
0.380 & 124 & 233 & 240 & 240 & 240 & 240 \\ 
0.436 & 124 & 285 & 315 & 317 & 317 & 318 \\ 
0.494 & 126 & 336 & 398 & 398 & 398 & 402 \\ 
0.555 & 128 & 386 & 485 & 486 & 486 & 490 \\ 
0.617 & 129 & 433 & 579 & 583 & 583 & 588 \\ 
0.682 & 131 & 477 & 626 & 702 & 697 & 703 \\ 
0.749 & 133 & 519 & 661 & 833 & 809 & 889 \\ 
0.818 & 134 & 557 & 692 & 859 & 850 & 978 \\ 
0.854 & 135 & 576 & 706 & 871 & 868 & 999 \\ 
0.926 & 137 & 611 & 733 & 893 & 903 & 1040 \\ 
1.0 & 139 & 644 & 758 & 914 & 935 & 1080 \\ 
\hline \hline
\end{tabular}} \hfill
    \begin{tabular}[t]{|c|c|c|c|c|c|c|c|}
\hline \hline
$D_{II}$ & $m(0^{-+})$ & $m(0^{++})$ & $m(1^{--})$ & $m(1^{+-})$ & $m(1^{++})$ & $\mathcal{M}$ \\ \hline \hline
0.03 & 25 & 25 & 25 & 25 & 25 & 18 \\ 
0.09 & 32 & 32 & 32 & 32 & 32 & 28 \\ 
0.15 & 49 & 49 & 49 & 49 & 49 & 47 \\ 
0.21 & 91 & 92 & 98 & 98 & 98 & 94 \\ 
0.27 & 132 & 189 & 198 & 198 & 198 & 194 \\ 
0.33 & 128 & 282 & 318 & 323 & 323 & 324 \\ 
0.39 & 128 & 357 & 463 & 465 & 461 & 460 \\ 
0.45 & 129 & 419 & 549 & 629 & 629 & 616 \\ 
0.51 & 131 & 470 & 588 & 734 & 721 & 787 \\ 
0.57 & 132 & 512 & 617 & 753 & 758 & 851 \\ 
0.63 & 133 & 547 & 641 & 769 & 788 & 911 \\ 
0.69 & 134 & 577 & 662 & 782 & 814 & 955 \\ 
0.75 & 135 & 602 & 682 & 794 & 837 & 993 \\ 
0.81 & 136 & 624 & 700 & 804 & 858 & 1028 \\ 
0.87 & 137 & 644 & 716 & 814 & 877 & 1061 \\ 
0.93 & 138 & 661 & 732 & 822 & 895 & 1091 \\ 
\hline \hline
\end{tabular} 
    \label{tab:model_I_and_II}
\end{table*}
\begin{table*}
    \centering
    \caption{Tabulated values of the meson masses extracted from the BSEs for model {III} set to a physical scale {(MeV)}  and the corresponding parameter $\mathcal{M}$.  
    {In some cases, indicated by a $-$ sign, we have not been able to 
    extract the on-shell masses of the first radial excitations within numerical accuracy. In principle those could be extracted using well-established extrapolation procedures in the eigenvalue curves. However, since these are not the main topic of this work and the general trend can already be seen from the available data, we did not invest the (considerable) CPU-time.}
    }
    \begin{tabular}{|c|c|c|c|c|c|c|c|c|c|c|c|}
\hline \hline
$D_{III}$ & $m(0^{-+})$ & $m(0^{++})$ & $m(1^{--})$ & $m(1^{+-})$ & $m(1^{++})$ & $m(0^{-+\prime})$ & $m(0^{++\prime})$ & $m(1^{--\prime})$ & $m(1^{+-\prime})$ & $m(1^{++\prime})$ & $ \mathcal{M}$ \\ \hline \hline
0.1 & 25 & 26 & 25 & 26 & 26 & - & - & 27 & - & - & 21 \\ 
0.15 & 38 & 39 & 38 & 40 & 40 & - & - & 42 & - & - & 39 \\ 
0.2 & 57 & 61 & 59 & 62 & 62 & 63 & 66 & 65 & 63 & 65 & 62 \\ 
0.25 & 79 & 91 & 87 & 95 & 95 & 95 & 99 & 96 & 99 & 99 & 99 \\ 
0.3 & 100 & 123 & 123 & 136 & 136 & 135 & 139 & 132 & 137 & 136 & 138 \\ 
0.35 & 115 & 153 & 165 & 186 & 184 & 183 & 188 & 187 & 189 & 188 & 187 \\ 
0.4 & 124 & 181 & 212 & 242 & 239 & 238 & 246 & 240 & 244 & 245 & 240 \\ 
0.45 & 130 & 211 & 263 & 303 & 299 & 299 & 306 & 301 & 307 & 307 & 300 \\ 
0.5 & 132 & 241 & 314 & 365 & 360 & 360 & - & 361 & - & - & 368 \\ 
0.55 & 134 & 271 & 364 & 426 & 419 & 421 & - & 424 & - & - & 430 \\ 
0.6 & 135 & 301 & 411 & 484 & 475 & 481 & - & 485 & - & - & 488 \\ 
0.65 & 137 & 331 & 456 & 538 & 529 & 537 & 551 & 543 & 551 & 550 & 540 \\ 
0.7 & 137 & 360 & 497 & 588 & 578 & 589 & - & 594 & - & - & 597 \\ 
0.75 & 138 & 387 & 535 & 633 & 623 & 638 & - & 647 & - & - & 650 \\ 
0.8 & 139 & 414 & 571 & 674 & 666 & 687 & - & 696 & - & - & 697 \\ 
0.85 & 139 & 438 & 604 & 710 & 705 & 731 & - & - & - & - & 736 \\ 
0.9 & 141 & 463 & 635 & 745 & 742 & 774 & 797 & 782 & 799 & 799 & 780 \\ 
0.95 & 143 & 487 & 666 & 778 & 778 & 815 & - & - & - & - & 820 \\ 
1 & 144 & 508 & 693 & 807 & 810 & 850 & - & 860 & - & - & 858 \\ 
\hline \hline
\end{tabular} 
    \label{tab:model_III}
\end{table*}

\newpage

\bibliographystyle{JHEP}
\bibliography{refs}
\end{document}